\documentclass
[preprint,showpacs,showkeys,prc,10pt,endfloats,onecolumn]{revtex4}%
\usepackage{amsfonts}
\usepackage{amsmath}
\usepackage{amssymb}
\usepackage{graphicx}
\usepackage[nooneline]{caption2}%
\begin{document}
\preprint{HEP/123-qed}
\title{Shape transition in the even-even Cerium isotopes }
\author{B. Mohammed-Azizi}
\affiliation{University of Bechar, Bechar, Algeria}
\author{D. E. Medjadi}
\affiliation{Ecole Normale Superieure, Kouba, Algiers, Algeria}
\author{A. Helmaoui}
\affiliation{University of Bechar, Bechar, Algeria}

\begin{abstract}
The deformation energy of the even-even nuclei of the Cerium isotopic chain is
investigated by means of the Macroscopic-Microscopic method with a
semiclassical shell correction. We consider axially symmetric shapes. Binding
energy and two neutron separation energy are also evaluated. For the sake of
clarity several important details of the calculations are also given. It turns
out that all these nuclei have prolate equilibrium shape. The regions of
maximum deformation are obtained around $N=64$ and $N=102$. There is no
critical-point of quantum phase transition in this isotopic chain.

\end{abstract}
\startpage{1}
\endpage{ }
\maketitle

\section{Introduction}

Nowadays it is well established that the majority of nuclei possess a nonzero
intrinsic electric quadrupole moment (IEQM). This feature means that the
charge distribution inside the nucleus deviates from the spherical symmetry.
In other words, apart from very few nuclei, the surface of the nucleus is
generally not spherical in its ground state. The intrinsic quadrupole electric
moments (or equivalently the nuclear deformation) can be deduced from two
types of measurements:

\textbullet\qquad The reduced electric quadrupole transition probability,
$B(E2)$ \cite{08}

\textbullet\qquad The static electric quadrupole moments of ground and excited
states, $Q$ \cite{09}

It turns out that in a number of cases, the two methods of measurement do not
systematically lead to the same values. Important discrepancies occur for
several nuclei. This is essentially due to the fact that not only different
experimental techniques are used but above all, because different models can
be implemented to deduce the nuclear deformation for the both cases. In
Ref.\cite{013} it is stated that deformations deduced from $B(E2)$ have a
"more general character". In other words,\textquotedblleft$B(E2)$%
-type\textquotedblright\ data reflect not only static nuclear deformation
(permanent deviation of the nuclear shape from sphericity), but also dynamic
deformation. Furthermore, $B(E2)$ measurements are model independent and thus
are generally more reliable. This is corroborated by the fact that the only
systematic compilation in which the deformation of the ground state is given
explicitly is based on $B(E2;0^{+}\rightarrow2^{+})$ and has been published in
Ref.\cite{08}. In the present work, experimental values refer to these ones.

Theoretical approaches to the deformation energy can be divided into two
categories; Dynamic calculations to find the shape of the ground state (or
even of excited states) and static calculations by determining the absolute
minimum (ground state) or multiple minima (shape isomers) in the potential
energy surface (PES) for a given nucleus. Thus, on the one hand, we have the
so-called collective models, which themselves are subdivided into two groups:
The "Geometric Collective Model" also called "the Collective Bohr Hamiltonian"
(CBH and its variants) and the "Algebraic Model", well known under the name of
the "Interacting Boson Model" (IBM and its variants) \cite{08x}. On the other
hand, "particle models" consider the nucleus as a collection of interacting
nucleons (fermions). In practice, the classical $N$-body problem can be
approximately solved by the usual approximation of the mean field with
eventually residual interactions. In this respect, the \textquotedblleft
best\textquotedblright\ mean field is deduced after applying a variational
principle in the Hartree-Fock-Bogoliubov method (HFB). In this model, the
determination of the potential energy surface (PES) of the nucleus amounts to
perform constrained Hartree-Fock-Bogoliubov (CHFB) calculations \cite{08xx}.
We will not address very complicated methods \textquotedblleft beyond the mean
field\textquotedblright\ such as the Quasiparticle Random Phase Approximation
(QRPA) or the Generator-Coordinate-Method (GCM) methods which are unsuitable
in practice for large scale calculations.\newline Because of CHFB calculations
are time consumers, especially in large studies, Microscopic-Macroscopic
method (Mic-Mac) constitutes a good alternative which, is up to now,
implemented \cite{08b}. In the present work, we use an improved variant of
this method. The word \textquotedblleft improved\textquotedblright\ means that
we use semi-classical method to avoid the well-known drawbacks (spurious
dependence on two mathematical parameters) of the standard Strutinsky shell
correction (see text below).

The present study is devoted to the deformation energy, equilibrium nuclear
shapes and binding energy of the ground state of the even-even cerium
isotopes. There are many reasons to this choice. One of them is to re-test our
previous calculations. In effect, similar calculations have been already
performed by us in the xenon, barium, and cerium region Ref \cite{014}.
However because the phenomenological mean potential varies smoothly with $N$
and $Z$, we have made, in the past, a rough approximation by choosing the same
set of parameters for the phenomenological mean potential, for the all treated
nuclei. Originally, this approximation was done only for simplifying the
calculations. Here, contrarily to that study, each nucleus has its
\textquotedblleft own\textquotedblright\ mean potential with a specific set of
parameters. In this way it is possible to evaluate in a rigorous way the
uncertainty introduced in the previous calculations. Apart from this remark,
there are several main other reasons which could justify this choice: (i)
First, it should be interesting to see how the deformation energy and binding
energy vary with the neutron number (N) for this isotopic chain. (ii) Second,
the present study extends the previous calculations to all cerium isotopes up
to the drip lines (34 versus 13 nuclei). (iii) Third, we also will attempt to
deduce, from potential energy surface (PES) curves, the shape transition from
spherical to axially deformed nuclei, looking for the so-called $X(5)$
critical-point between U(5) and SU(3) symmetry limits of the IBM
\cite{014b}-\cite{014bb} .

It is worth to recall briefly some information deduced from the literature for
the cerium isotopes. In the past, a number of experimental as well as
theoretical studies have been done for the cerium isotopes. \ Among the
numerous studies, we only cite some of them: In 2005 Smith et al \cite{01}
have studied excited states of $^{122}Ce$ up to spin $14\hbar$ deducing a
probable quadrupole deformation of about $\beta\approx0.35$. The deformed
nucleus $^{130}Ce$ has been studied in 1985, using the techniques of in-beam
gamma -ray spectroscopy \cite{02}. The corresponding data have been
interpreted in terms of the cranking model by assuming a prolate deformation
with $\varepsilon_{2}\approx0.25$ ($\beta\approx0.27$). High-spin states in
$^{132}Ce$ have been also studied by A.J. Kirwan et al. \cite{03}. They found
a superdeformed band with deformation $\beta\approx0.4$ much more larger that
the ground state deformation $\left(  \beta\approx0.2\right)  $. E.
Michelakakis et al \cite{04} by evaluating $\gamma-$ray transitions in
$^{142}Ce$ and $^{144}Ce$ conclude that in cerium isotopes (near the
beta-stable line) the onset of nuclear deformation occur between $N=86$ and
$N=88$. "Pure" theoretical calculations have been performed in Ref.\cite{05}
and \cite{06} with projected shell model (PSM) and Hartree-Bogoliubov ansatz
in the valence space respectively for $^{122}Ce$ and $^{124-132}Ce$ for low
lying yrast spectra. Good values of energy levels and reduced transition
probabilities $B(E2,0^{+}\rightarrow2^{+})$ have been obtained respectively in
these two papers. Other approaches for the rich-neutron cerium isotopes have
been made in Ref.\cite{07}. A study of the shape transition from spherical to
axially deformed nuclei in the even Ce isotopes has been done in Ref
\cite{07b} using the nucleon-pair approximation of the shell model. The result
of a such study is that the transition has been found too rapid. Relativistic
Hartree-Fock-Bogoliubov theory has been used to predict ordinary halo for
$^{186}Ce,^{188}Ce,^{190}Ce,$ and giant halo for $^{192}Ce,^{194}%
Ce,^{196}Ce,^{198}Ce$ near the neutron drip line. Systematic studies about
nuclear deformations and masses of the ground state can be found in
Ref.\cite{010}-\cite{012}\ with respectively, the Finite-Range Droplet-Model
(FRDM) , Hartree-Fock-Bogoliubov (HFB), HFB+5-dimensional collective quadupole
Hamiltonian and Relativistic Mean Field (RMF) models.

\section{The Macroscopic-Microscopic method}

\subsection{Liquid drop model and microscopic corrections}

This method combines the so-called semi-empirical mass formula (or liquid drop
model) with shell and pairing corrections deduced from microscopic model. Thus
the binding energy is given as a function of \ nucleon numbers and deformation
parameter $($referred to as $\beta)$ by mean of the usual symbols:%

\begin{equation}
B(A,Z,\beta)=E_{LDM}(\beta)-\delta B_{micro}(\beta) \label{binding}%
\end{equation}
$\delta B_{micro}$ contains the shell and pairing correction (see text below).
The minus sign before $\delta B_{micro}$ is consistent with the convention
that the binding energy is defined as positive here. For the liquid drop model
we take the old version of Myers and Swiatecki \cite{015} (because of its
simplicity compared to more recent formulae). Here, there is no need to\ look
for very high accuracy in binding energy, because this is not the purpose of
the present work.%

\begin{equation}
E_{LDM}(\beta)=C_{V}A-C_{S}A^{2/3}B_{S}(\beta)-C_{C}Z^{2}A^{-1/3}B_{C}%
(\beta)+\varepsilon a_{pair}A^{-1/2}+C_{d}Z^{2}A^{-1} \label{ldm}%
\end{equation}
In Eq.(\ref{ldm}), we have the usual contributions of volume, surface and
coulomb energies. The different constants of Myers and Swiatecki are given in
appendix A. The shape dependence ($\beta$) of the surface and coulomb energies
are contained in $B_{S}(\beta)$ and $B_{C}(\beta)$. They are normalized to the
unity for a spherical nuclear surface. The latter is symbolized by $\beta=0.$
The two last terms in Eq.(\ref{ldm}) are respectively due to the smooth part
of the pairing energy and the correction of the Coulomb energy to account for
the diffuseness of the nucleus surface. The different constants will be fixed
later. The potential energy surface (PES without zero point energy correction)
is defined as follows:%

\begin{equation}
E_{PES}(\beta)=E_{LDM}(0)-B(A,Z,\beta)=\Delta E_{LDM}(\beta)+\delta
B_{micro}(\beta) \label{pes}%
\end{equation}
in which%
\begin{equation}
\Delta E_{LDM}(\beta)=E_{LDM}(0)-E_{LDM}(\beta)=C_{S}A^{2/3}\left[
B_{S}(\beta)-B_{S}(0)\right]  +C_{C}Z^{2}A^{-1/3}\left[  B_{C}(\beta
)-B_{C}(0)\right]  \label{dldm}%
\end{equation}
Constants $C_{V}$ and $C_{S}$ are expressed by means of three other constants
$a_{V},a_{S},$ and $\kappa$. For spherical shape, as said before, the
normalization is expressed by: $B_{S}(0)=B_{C}(0)=1.$ As it can be easily
seen, the potential energy surface is related only to two macroscopic
constants $C_{S}$ (which depends actually on $a_{S}$ and $\kappa$) and $C_{C}%
$. To calculate microscopic shell an pairing corrections contained in $\delta
B_{micro}$, we have to proceed in two steps. The first consists in solving the
Schrodinger equation and the second in deducing the shell and pairing
corrections in an appropriate way, as explained in the following.

\subsection{Microscopic model}

We briefly present the microscopic model which is based on the Schrodinger
equation of the deformed independent particle model:%
\begin{equation}
\hat{H}(\beta)\mid\Psi_{i}(\beta)\rangle=\varepsilon_{i}(\beta)\mid\Psi
_{i}(\beta)\rangle\label{Schro}%
\end{equation}
where $\mid\Psi_{i}\rangle$ and $\varepsilon_{i}$ are respectively the
eigenfunctions and the associated eigenvalues of nucleons. Hamiltonian
$\hat{H}$ contains four contributions which are: (i) kinetic energy, (ii)
central deformed mean field, \ (iii) spin-orbit and (iv) Coulomb interactions.
We perform analogous calculations as in Nilsson model but our deformed mean
potential is of Woods-Saxon type and therefore is "more realistic". Although
calculations are not self consistent, they are microscopic. It is to be noted
that our Schrodinger equation has a form which is very close to the one of the
Skyrme-Hartree-Fock method.\newline Eq.(\ref{Schro}) is solved by our FORTRAN
program described in details in Ref \cite{012a} and improved in two successive
versions \cite{012b} and \cite{012c}.

\subsection{Microscopic corrections}

Microscopic corrections are defined as the sum of shell and pairing
corrections which themselves are calculated separately for each kind of nucleons.%

\begin{equation}
\delta B_{micro}(\beta)=\delta E_{shell}(N,\beta)+\delta E_{shell}%
(Z,\beta)+\delta P_{pairing}(N,\beta)+\delta P_{pairing}(Z,\beta)
\label{deltab}%
\end{equation}
In this formula the shell correction is defined by the usual Strutinsky
prescription, i.e., as the difference between the sum of the single particle
energies (which contains the shell effects) and an averaged (or smoothed) sum
(which is free from shell effects)%

\begin{equation}
\delta E_{shell}(N\text{ }or\text{ }Z)=%
%TCIMACRO{\dsum \limits_{i=1}^{NorZ}}%
%BeginExpansion
{\displaystyle\sum\limits_{i=1}^{NorZ}}
%EndExpansion
\varepsilon_{i}(\beta)-\overline{\sum_{i=1}\varepsilon_{i}\left(
\beta\right)  } \label{shell correction}%
\end{equation}
Energies $\varepsilon_{i}(\beta)$ are deduced from Eq.(\ref{Schro}). In our
procedure, the second sum is found by means of a semi-classical way instead a
Strutinsky smoothing procedure, see Ref.\cite{012d} . This avoids the
well-known weakness of the standard shell correction method, namely, the
dependence on two unphysical parameters which are the "smoothing" parameter
and the order of the curvature correction. Moreover, it has been clearly shown
that Strutinsky level density method is only an approximation of that of the
semi-classical theory \cite{012e}.\newline The "pure" pairing correlation
energy is defined by:%
\[
P(\beta)=%
%TCIMACRO{\dsum \limits_{i=1}^{\infty}}%
%BeginExpansion
{\displaystyle\sum\limits_{i=1}^{\infty}}
%EndExpansion
2\varepsilon_{i}(\beta)\upsilon_{i}^{2}-%
%TCIMACRO{\dsum \limits_{i=1}^{N/2orZ/2}}%
%BeginExpansion
{\displaystyle\sum\limits_{i=1}^{N/2orZ/2}}
%EndExpansion
2\varepsilon_{i}(\beta)-\frac{\Delta^{2}}{G}%
\]
where $\upsilon_{i}^{2}$, $\Delta$ and $\lambda$ are the usual occupation
probabilities, gap and Fermi energy of the BCS approximation (the factor "2"
is simply due to the Kramers degeneracy). Since the smooth part of pairing
correlations is already contained in the liquid drop model, we have to add
only the one due to the shell oscillations of the level density. This
contribution is defined by means of a similar formula to
Eq.(\ref{shell correction})
\begin{equation}
\delta P_{pairing}(N\text{ }or\text{ }Z,\text{ }\beta)=P(\beta)-\overline
{P(\beta)} \label{pairing correction}%
\end{equation}
where the averaged pairing is defined as $\overline{P(\beta)}=(1/2)g_{semicl.}%
(\lambda)\overline{\Delta}^{2}$. We use a simple BCS method to account for
pairing correlations. To calculate Eq (\ref{shell correction}) and
(\ref{pairing correction}) we follow the method detailed in Ref.\cite{012d}
with its FORTRAN code. The treatment of the pairing has also been explained in
Ref.\cite{014} and references quoted therein.

\subsection{Numerical constants and prescriptions}

\subsubsection{Constants of the Microscopic model}

For each kind of particles the mean central \ and the mean spin-orbit field
are written as \cite{012a}:%

\begin{equation}
V(\beta)=\frac{V_{0}}{1+\exp(R_{V}L_{V}(\beta)/a_{0})}%
\ \ \ \ \ \ \ \ \ \ \ V_{SO}(\beta)=\lambda\left(  \frac{\hbar}{2Mc}\right)
\frac{V_{0}}{1+\exp(R_{SO}L_{SO}(\beta)/a_{0})} \label{pot}%
\end{equation}
where $L_{V}(\beta)$ and $L_{SO}(\beta)$ contains the information on the
deformation. In fact, these functions contain 9 constants: $V_{0neut},$
$V_{0prot},$ $R_{Vneut},$ $R_{Vprot},$ $R_{SO-neut},$ $R_{SO-prot},$ $a_{0},$
$\lambda_{neut},$ $\lambda_{prot}$. These quantities are taken from the
"universal" parameters \cite{015b} (see appendix B) which is an optimized
set.\newline The Coulomb mean field is approximated by a uniform charge
distribution inside a deformed surface. The volume conservation is therefore
$Vol=(4/3)\pi R_{ch}^{3}$with the simple assumption $R_{ch}=R_{Vprot}.$

\subsubsection{Constants of the liquid drop model}

As already stated, we have chosen the parameters of Myers and Swiatecki (see
table \ref{ms2}) because this set contains a reduced number of parameters with
respect to more modern formulae.

\begin{table}[ptbh]
$\ $%
\begin{tabular}
[c]{c|cccccc}
& $a_{V}$ & $a_{S}$ & $C_{C}$ & $\kappa$ & $C_{d}$ & $a_{pair}$\\\hline
Myers and Swiatecki & $15.67MeV$ & $18.56MeV$ & $0.72MeV$ & $1.79$ & $1.21MeV$
& $11MeV$%
\end{tabular}
\caption{Parameters of the liquid drop model in the Myers and Swiatecki
version \cite{015}}%
\label{ms2}%
\end{table}

All the constants are needed in the binding energy whereas only $a_{S}%
,C_{C},\kappa$ play a role in the potential energy surface.

\subsubsection{Nuclear mass excesses}

Nuclear masses are deduced as mass excesses:\newline$M_{excess}(A,Z)=ZM_{H}%
+(A-Z)M_{n}-B(A,Z)$\newline where $M_{H}=7.289034MeV$ is the hydrogen mass
excess and $M_{n}=8.071431MeV$ the neutron mass excess. This makes comparisons
with experimental values easiest.

\section{Results}

In our previous paper \cite{014}\ calculations for isotopes $^{116-130}Ce$
showed that the equilibrium deformations ($\beta\approx0.25-0.30$) have always
been obtained for symmetric prolate shapes ($\gamma=0%
%TCIMACRO{\U{b0}}%
%BeginExpansion
{{}^\circ}%
%EndExpansion
$). Results obtained in Ref.\cite{018} with a similar approach for the nuclei
$^{116-130}Ce,$ corroborate this fact. For these reasons, we think that it is
needless to account for the axial asymmetry in a "pure" static study of the
equilibrium deformation. However, we have to consider prolate ($\gamma=0%
%TCIMACRO{\U{b0}}%
%BeginExpansion
{{}^\circ}%
%EndExpansion
$) as well as oblate ($\gamma=60%
%TCIMACRO{\U{b0}}%
%BeginExpansion
{{}^\circ}%
%EndExpansion
$) nuclear shapes. In this regard, it is worth remembering that oblate shape
given by ($\beta>0,\gamma=60%
%TCIMACRO{\U{b0}}%
%BeginExpansion
{{}^\circ}%
%EndExpansion
$) is equivalent to the set ($\beta<0,\gamma=0%
%TCIMACRO{\U{b0}}%
%BeginExpansion
{{}^\circ}%
%EndExpansion
$).

\subsection{Comparison between the different contributions entering in the
potential energy surface}

\begin{figure}[ptb]
\includegraphics[width=160mm,keepaspectratio]{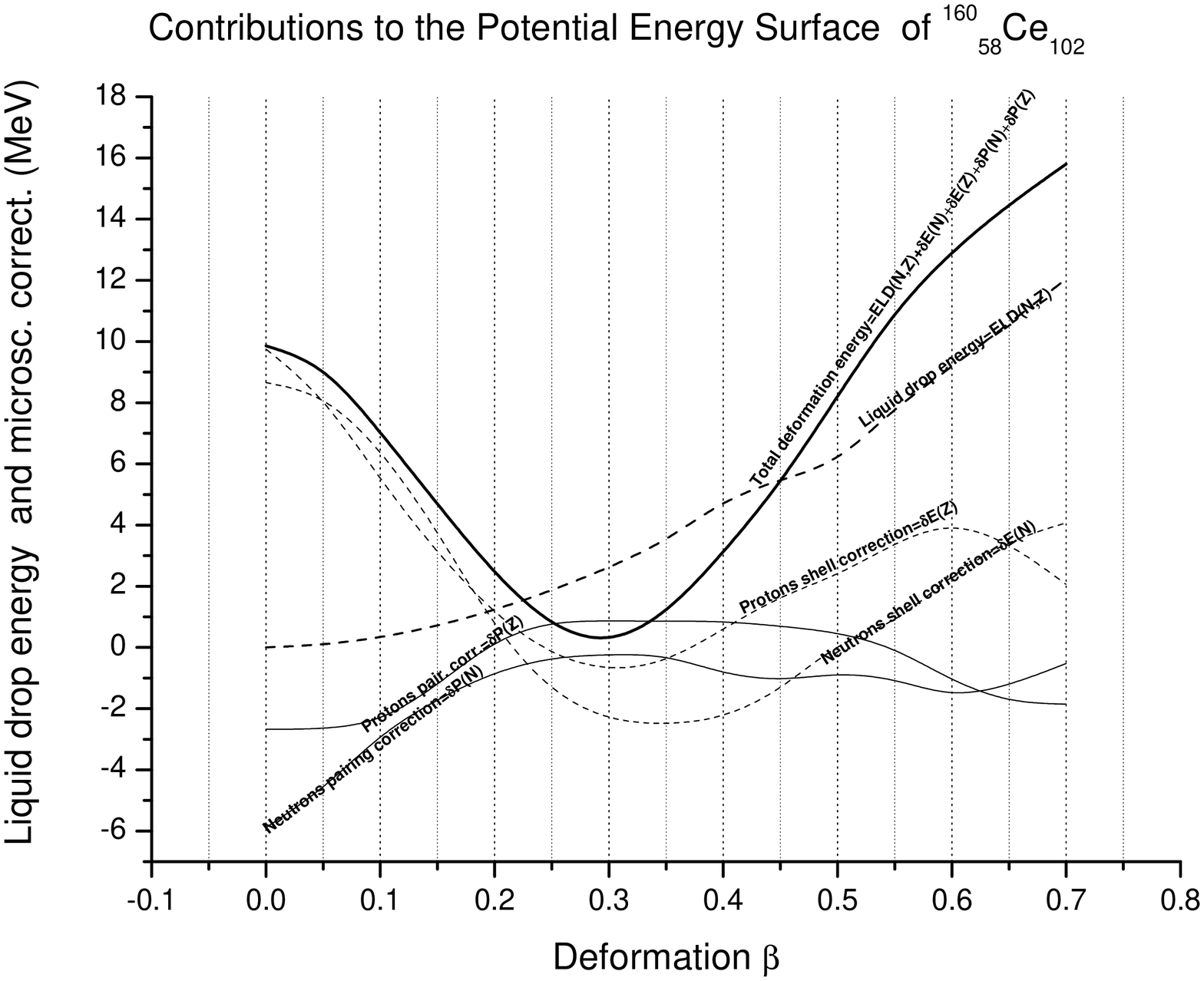}\caption{Contributions
of the shell and pairing corrections for the two kind of nucleons and the one
of the liquid drop model to the total potential energy surface of the nucleus
$^{160}Ce.$}%
\label{contri}%
\end{figure}

It could be useful to compare the importance of the different terms entering
in the right hand side of Eq.(\ref{deltab}). In this respect, we have drawn in
Fig.\ref{contri} for axially prolate shape, the four microscopic contributions
$\delta E_{shell}(N,\beta)$, $\delta E_{shell}(Z,\beta)$, $\delta
P_{pairing}(N,\beta)$, $\delta P_{pairing}(Z,\beta)$ for the case of
$^{160}Ce$ as functions of $\beta$. Following the cited order, we can say that
the difference between the highest and lowest values (in the interval
$\beta\in\lbrack0.0,0.7]$) are respectively about
$11.0MeV,10.5MeV,5.7MeV,3.5MeV$ for the four corrections. Thus, these
variations show that the shell corrections $\delta E_{shell}(N,\beta)$,
$\delta E_{shell}(Z,\beta)$ are more important than $\delta P_{pairing}%
(N,\beta)$, $\delta P_{pairing}(Z,\beta)$ and have a clear minimum at
respectively $\beta=0.35$ and $\beta=0.30.$ It is well known that for each
kind of nucleon the shell correction is in opposite phase with respect to the
pairing correction (this means for that when $\delta E_{shell}(N,\beta)$
increases with $\beta$, $\delta P_{pairing}(N,\beta)$ decreases and vice versa).

\begin{figure}[ptb]
\includegraphics[width=140mm,keepaspectratio]{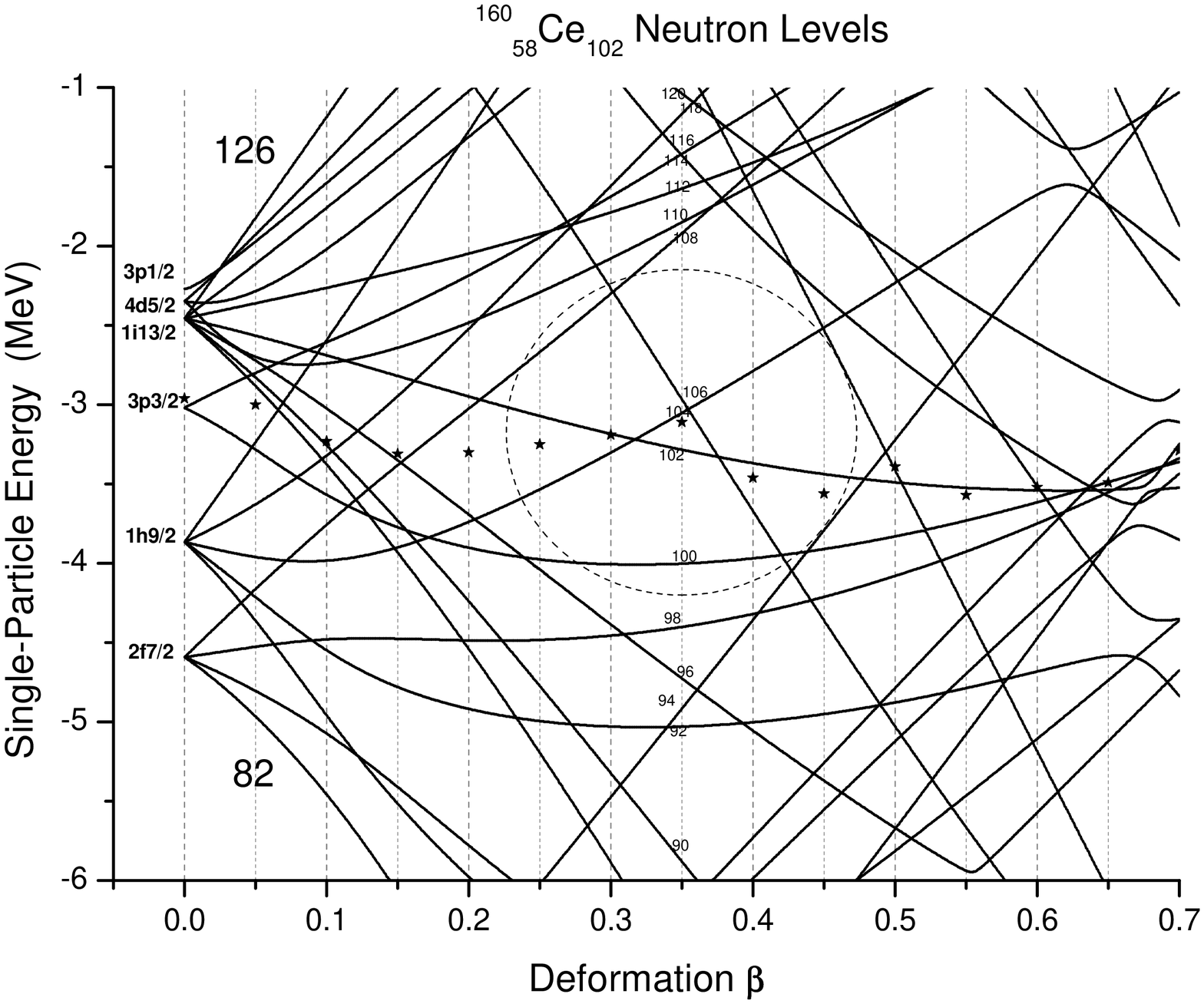}\caption{Single-particle
energies of the microscopic model as function of deformation for prolate
shapes $\left(  \beta>0\right)  $ for the nucleus $^{160}Ce$. Spherical
spectroscopic notation is given for spherical deformation $\left(
\beta=0\right)  .$The circle in dotted line indicates the area of lowest level
density.}%
\label{spe}%
\end{figure}

Contrarily to these curves, the liquid drop model is strictly increasing with
$\beta$, and its minimum occurs always at the beginning $\beta=0.$(spherical
shape). When all the contributions are added, the minimum of the potential
energy surface of the nucleus is reached at about $\beta=0.3$ and is mainly
due to the shell corrections. When $\beta$ becomes more and more, larger the
contribution of the liquid drop energy becomes preponderant so that the
equilibrium deformation occurs generally between $\beta=0$ and $\beta=0.4$.
Because of the convention of the sign stated before, $\delta B_{micro}$
defined in Eq.(\ref{binding}) must be negative in order to increase the
binding energy of the nucleus. Since the shell corrections (for protons and
neutrons) play a major role in $\delta B_{micro}$, it is naturally expected
that negative (but absolute large) values of shell correction contribute to
increase the binding energy of the nucleus. In this respect, it is well known
that the shell correction is essentially determined by the distribution of
single-particle levels in the vicinity of the sharp Fermi level (defined here
as the midway between the last occupied level and the first empty level).
Following Ref.\cite{017}, we can state that "the nuclear ground state, as well
as any other relatively stable state, should correspond to the lowest possible
degeneracy, or, in other words, the lowest density of state near the Fermi
level". This is illustrated in Fig. \ref{spe} where the single-particle levels
are drawn as function of the deformation $\beta$ ($\gamma$ being fixed at
$\gamma=0%
%TCIMACRO{\U{b0}}%
%BeginExpansion
{{}^\circ}%
%EndExpansion
$). To this end we have used the FORTRAN code of Ref.\cite{012a} and
\cite{012c}. The area where the single-particle level density is low near the
Fermi level (black stars) is indicated by a circle. Thus, it is not so
surprising that, it is in this region where the neutron shell correction
becomes the most important, involving a minimum in the PES of the nucleus.

\subsection{Equilibrium deformations}

They are given in table (\ref{t2}) for prolate as well as oblate shapes (see
table legend for details). The minima of PES for the corresponding wells are
denoted $minpro$ and $minobl$. The deformation energy is defined as the
difference $E_{def}=E_{PES}(0)-E_{PES}^{\min}(\beta)$, i.e., the difference
between the potential energy surface for a spherical shape and the one
corresponding to the absolute minimum of PES. Permanent deformations will be
in principle characterized by large values of $E_{def}$ and are responsible of
rotational spectra.

From this table, some remarks may be drawn:\newline(i) Two regions of prolate
deformation are found. They occur around $N=64$ and $N=102$ with maximum
deformation about $\beta\approx0.30.$ The deformation energy (between
spherical and deformed shape) is about $6.70MeV$ for $N=64$ and $9.30MeV$ for
$N=102$ and decreases from either side from these two nuclei.\newline(ii)
Spherical deformation occur at and near the (magic) numbers $N=82$ and $N=128$
(not shown)$.$\newline(iii) The deformation energy decreases from $N=64$
(maximum) to $N=82$ (minimum) and reincreases again to $N=102$ (maximum). We
have found graphically that the first inflexion point occurs between $N=72$
and $N=74$ and a second inflexion point is found between $N=90$ and $N=92$.
One can consider (somewhat arbitrarily) that spherical shapes occur
approximately between these two limits.\newline(iv) The minima of prolate
equilibrium deformations are, by far, always deeper compared to the ones of
the oblate minima ( $minpro\ll$ \ $minobl$). In other words cerium isotopes
prefer, by far, prolate shapes. In other words, the deformation energy
increases in average with the asymmetry $\gamma.$ This justifies a posteriori
that, in a static study of the equilibrium deformation, it is needless to
account for axial asymmetry. It is worth to remember that most of nuclei of
the chart have prolate shape (see Ref.\cite{012cc}).\newline(iv) Even though
the experimental deformations are known only in absolute value from $B(E2),$ a
good agreement is obtained if one excepts the three "nearly magic" nuclei
$^{138-142}Ce$.

\begin{table}[ptbh]%
\begin{tabular}
[c]{llllllll|llllllll}%
$N$ & $A$ & $\beta_{pro}$ & $minpro$ & $\beta_{obl}$ & $minobl$ & $E_{def}$ &
$\left\vert \beta_{\exp}\right\vert $ & $N$ & $A$ & $\beta_{pro}$ & $minpro$ &
$\beta_{obl}$ & $minobl$ & $E_{def}$ & $\left\vert \beta_{\exp}\right\vert $\\
\multicolumn{1}{c}{} & \multicolumn{1}{c}{} & \multicolumn{1}{c}{} &
\multicolumn{1}{c}{$\left(  MeV\right)  $} & \multicolumn{1}{c}{} &
\multicolumn{1}{c}{$\left(  MeV\right)  $} & \multicolumn{1}{c}{$\left(
MeV\right)  $} & \multicolumn{1}{c|}{} & \multicolumn{1}{|c}{} &
\multicolumn{1}{c}{} & \multicolumn{1}{c}{} & \multicolumn{1}{c}{$\left(
MeV\right)  $} & \multicolumn{1}{c}{} & \multicolumn{1}{c}{$\left(
MeV\right)  $} & \multicolumn{1}{c}{$\left(  MeV\right)  $} &
\multicolumn{1}{c}{}\\\hline\hline
\multicolumn{1}{c}{$58$} & \multicolumn{1}{c}{$116$} &
\multicolumn{1}{c}{$0.30$} & \multicolumn{1}{c}{$0.90$} &
\multicolumn{1}{c}{$-0.21$} & \multicolumn{1}{c}{$3.62$} &
\multicolumn{1}{c}{$4.80$} & \multicolumn{1}{c|}{} & \multicolumn{1}{|c}{$92$}
& \multicolumn{1}{c}{$150$} & \multicolumn{1}{c}{$0.25$} &
\multicolumn{1}{c}{$1.23$} & \multicolumn{1}{c}{$-0.17$} &
\multicolumn{1}{c}{$4.45$} & \multicolumn{1}{c}{$5.12$} &
\multicolumn{1}{c}{$0.31$}\\
\multicolumn{1}{c}{$60$} & \multicolumn{1}{c}{$118$} &
\multicolumn{1}{c}{$0.32$} & \multicolumn{1}{c}{$0.88$} &
\multicolumn{1}{c}{$-0.23$} & \multicolumn{1}{c}{$4.07$} &
\multicolumn{1}{c}{$5.87$} & \multicolumn{1}{c|}{} & \multicolumn{1}{|c}{$94$}
& \multicolumn{1}{c}{$152$} & \multicolumn{1}{c}{$0.27$} &
\multicolumn{1}{c}{$1.21$} & \multicolumn{1}{c}{$-0.19$} &
\multicolumn{1}{c}{$5.05$} & \multicolumn{1}{c}{$6.40$} & \multicolumn{1}{c}{}%
\\
\multicolumn{1}{c}{$62$} & \multicolumn{1}{c}{$120$} &
\multicolumn{1}{c}{$0.32$} & \multicolumn{1}{c}{$1.03$} &
\multicolumn{1}{c}{$-0.23$} & \multicolumn{1}{c}{$4.33$} &
\multicolumn{1}{c}{$6.19$} & \multicolumn{1}{c|}{} & \multicolumn{1}{|c}{$96$}
& \multicolumn{1}{c}{$154$} & \multicolumn{1}{c}{$0.28$} &
\multicolumn{1}{c}{$0.64$} & \multicolumn{1}{c}{$-0.21$} &
\multicolumn{1}{c}{$4.94$} & \multicolumn{1}{c}{$7.47$} & \multicolumn{1}{c}{}%
\\
\multicolumn{1}{c}{$64$} & \multicolumn{1}{c}{$122$} &
\multicolumn{1}{c}{$0.31$} & \multicolumn{1}{c}{$1.16$} &
\multicolumn{1}{c}{$-0.23$} & \multicolumn{1}{c}{$4.23$} &
\multicolumn{1}{c}{$6.68$} & \multicolumn{1}{c|}{} & \multicolumn{1}{|c}{$98$}
& \multicolumn{1}{c}{$156$} & \multicolumn{1}{c}{$0.29$} &
\multicolumn{1}{c}{$0.66$} & \multicolumn{1}{c}{$-0.22$} &
\multicolumn{1}{c}{$5.13$} & \multicolumn{1}{c}{$8.44$} & \multicolumn{1}{c}{}%
\\
\multicolumn{1}{c}{$66$} & \multicolumn{1}{c}{$124$} &
\multicolumn{1}{c}{$0.30$} & \multicolumn{1}{c}{$1.47$} &
\multicolumn{1}{c}{$-0.21$} & \multicolumn{1}{c}{$4.15$} &
\multicolumn{1}{c}{$6.17$} & \multicolumn{1}{c|}{$0.30$} &
\multicolumn{1}{|c}{$100$} & \multicolumn{1}{c}{$158$} &
\multicolumn{1}{c}{$0.29$} & \multicolumn{1}{c}{$0.71$} &
\multicolumn{1}{c}{$-0.22$} & \multicolumn{1}{c}{$5.14$} &
\multicolumn{1}{c}{$9.08$} & \multicolumn{1}{c}{}\\
\multicolumn{1}{c}{$68$} & \multicolumn{1}{c}{$126$} &
\multicolumn{1}{c}{$0.29$} & \multicolumn{1}{c}{$1.75$} &
\multicolumn{1}{c}{$-0.21$} & \multicolumn{1}{c}{$3.87$} &
\multicolumn{1}{c}{$5.43$} & \multicolumn{1}{c|}{$0.33$} &
\multicolumn{1}{|c}{$102$} & \multicolumn{1}{c}{$160$} &
\multicolumn{1}{c}{$0.30$} & \multicolumn{1}{c}{$0.32$} &
\multicolumn{1}{c}{$-0.22$} & \multicolumn{1}{c}{$4.52$} &
\multicolumn{1}{c}{$9.27$} & \multicolumn{1}{c}{}\\
\multicolumn{1}{c}{$70$} & \multicolumn{1}{c}{$128$} &
\multicolumn{1}{c}{$0.27$} & \multicolumn{1}{c}{$1.82$} &
\multicolumn{1}{c}{$-0.21$} & \multicolumn{1}{c}{$3.48$} &
\multicolumn{1}{c}{$4.67$} & \multicolumn{1}{c|}{$0.29$} &
\multicolumn{1}{|c}{$104$} & \multicolumn{1}{c}{$162$} &
\multicolumn{1}{c}{$0.29$} & \multicolumn{1}{c}{$0.71$} &
\multicolumn{1}{c}{$-0.22$} & \multicolumn{1}{c}{$4.42$} &
\multicolumn{1}{c}{$9.08$} & \multicolumn{1}{c}{}\\
\multicolumn{1}{c}{$72$} & \multicolumn{1}{c}{$130$} &
\multicolumn{1}{c}{$0.25$} & \multicolumn{1}{c}{$2.02$} &
\multicolumn{1}{c}{$-0.2$} & \multicolumn{1}{c}{$3.27$} &
\multicolumn{1}{c}{$3.34$} & \multicolumn{1}{c|}{$0.26$} &
\multicolumn{1}{|c}{$106$} & \multicolumn{1}{c}{$164$} &
\multicolumn{1}{c}{$0.29$} & \multicolumn{1}{c}{$1.00$} &
\multicolumn{1}{c}{$-0.23$} & \multicolumn{1}{c}{$4.23$} &
\multicolumn{1}{c}{$8.44$} & \multicolumn{1}{c}{}\\
\multicolumn{1}{c}{$74$} & \multicolumn{1}{c}{$132$} &
\multicolumn{1}{c}{$0.20$} & \multicolumn{1}{c}{$1.90$} &
\multicolumn{1}{c}{$-0.17$} & \multicolumn{1}{c}{$2.60$} &
\multicolumn{1}{c}{$1.97$} & \multicolumn{1}{c|}{$0.26$} &
\multicolumn{1}{|c}{$108$} & \multicolumn{1}{c}{$166$} &
\multicolumn{1}{c}{$0.28$} & \multicolumn{1}{c}{$1.16$} &
\multicolumn{1}{c}{$-0.23$} & \multicolumn{1}{c}{$3.92$} &
\multicolumn{1}{c}{$7.57$} & \multicolumn{1}{c}{}\\
\multicolumn{1}{c}{$76$} & \multicolumn{1}{c}{$134$} &
\multicolumn{1}{c}{$0.16$} & \multicolumn{1}{c}{$1.28$} &
\multicolumn{1}{c}{$-0.14$} & \multicolumn{1}{c}{$1.63$} &
\multicolumn{1}{c}{$0.93$} & \multicolumn{1}{c|}{$0.19$} &
\multicolumn{1}{|c}{$110$} & \multicolumn{1}{c}{$168$} &
\multicolumn{1}{c}{$0.27$} & \multicolumn{1}{c}{$1.46$} &
\multicolumn{1}{c}{$-0.21$} & \multicolumn{1}{c}{$3.84$} &
\multicolumn{1}{c}{$6.39$} & \multicolumn{1}{c}{}\\
\multicolumn{1}{c}{$78$} & \multicolumn{1}{c}{$136$} &
\multicolumn{1}{c}{$0.10$} & \multicolumn{1}{c}{$0.04$} &
\multicolumn{1}{c}{$-0.07$} & \multicolumn{1}{c}{$0.18$} &
\multicolumn{1}{c}{$0.19$} & \multicolumn{1}{c|}{$0.17$} &
\multicolumn{1}{|c}{$112$} & \multicolumn{1}{c}{$170$} &
\multicolumn{1}{c}{$0.25$} & \multicolumn{1}{c}{$1.68$} &
\multicolumn{1}{c}{$-0.20$} & \multicolumn{1}{c}{$3.55$} &
\multicolumn{1}{c}{$5.33$} & \multicolumn{1}{c}{}\\
\multicolumn{1}{c}{$80$} & \multicolumn{1}{c}{$138$} &
\multicolumn{1}{c}{$0.00$} & \multicolumn{1}{c}{$-1.93$} &
\multicolumn{1}{c}{$0.00$} & \multicolumn{1}{c}{$-1.93$} &
\multicolumn{1}{c}{$0.00$} & \multicolumn{1}{c|}{$0.13$} &
\multicolumn{1}{|c}{$114$} & \multicolumn{1}{c}{$172$} &
\multicolumn{1}{c}{$0.25$} & \multicolumn{1}{c}{$1.97$} &
\multicolumn{1}{c}{$-0.19$} & \multicolumn{1}{c}{$3.20$} &
\multicolumn{1}{c}{$4.19$} & \multicolumn{1}{c}{}\\
\multicolumn{1}{c}{$82$} & \multicolumn{1}{c}{$140$} &
\multicolumn{1}{c}{$0.00$} & \multicolumn{1}{c}{$-3.96$} &
\multicolumn{1}{c}{$0.00$} & \multicolumn{1}{c}{$-3.96$} &
\multicolumn{1}{c}{$0.00$} & \multicolumn{1}{c|}{$0.10$} &
\multicolumn{1}{|c}{$116$} & \multicolumn{1}{c}{$174$} &
\multicolumn{1}{c}{$0.2$} & \multicolumn{1}{c}{$1.93$} &
\multicolumn{1}{c}{$-0.17$} & \multicolumn{1}{c}{$2.79$} &
\multicolumn{1}{c}{$2.95$} & \multicolumn{1}{c}{}\\
\multicolumn{1}{c}{$84$} & \multicolumn{1}{c}{$142$} &
\multicolumn{1}{c}{$0.00$} & \multicolumn{1}{c}{$-2.07$} &
\multicolumn{1}{c}{$0.00$} & \multicolumn{1}{c}{$-2.07$} &
\multicolumn{1}{c}{$0.00$} & \multicolumn{1}{c|}{$0.12$} &
\multicolumn{1}{|c}{$118$} & \multicolumn{1}{c}{$176$} &
\multicolumn{1}{c}{$0.17$} & \multicolumn{1}{c}{$1.71$} &
\multicolumn{1}{c}{$-0.16$} & \multicolumn{1}{c}{$2.17$} &
\multicolumn{1}{c}{$1.68$} & \multicolumn{1}{c}{}\\
\multicolumn{1}{c}{$86$} & \multicolumn{1}{c}{$144$} &
\multicolumn{1}{c}{$0.15$} & \multicolumn{1}{c}{$0.02$} &
\multicolumn{1}{c}{$-0.06$} & \multicolumn{1}{c}{$0.53$} &
\multicolumn{1}{c}{$0.50$} & \multicolumn{1}{c|}{$0.17$} &
\multicolumn{1}{|c}{$120$} & \multicolumn{1}{c}{$178$} &
\multicolumn{1}{c}{$0.14$} & \multicolumn{1}{c}{$1.39$} &
\multicolumn{1}{c}{$-0.14$} & \multicolumn{1}{c}{$1.60$} &
\multicolumn{1}{c}{$0.55$} & \multicolumn{1}{c}{}\\
\multicolumn{1}{c}{$88$} & \multicolumn{1}{c}{$146$} &
\multicolumn{1}{c}{$0.19$} & \multicolumn{1}{c}{$0.73$} &
\multicolumn{1}{c}{$-0.11$} & \multicolumn{1}{c}{$2.43$} &
\multicolumn{1}{c}{$1.99$} & \multicolumn{1}{c|}{$0.17$} &
\multicolumn{1}{|c}{$122$} & \multicolumn{1}{c}{$180$} &
\multicolumn{1}{c}{$0.0$} & \multicolumn{1}{c}{$0.3$} &
\multicolumn{1}{c}{$0.00$} & \multicolumn{1}{c}{$0.30$} &
\multicolumn{1}{c}{$-0.15$} & \multicolumn{1}{c}{}\\
\multicolumn{1}{c}{$90$} & \multicolumn{1}{c}{$148$} &
\multicolumn{1}{c}{$0.23$} & \multicolumn{1}{c}{$1.15$} &
\multicolumn{1}{c}{$-0.14$} & \multicolumn{1}{c}{$3.76$} &
\multicolumn{1}{c}{$3.15$} & \multicolumn{1}{c|}{$0.25$} &
\multicolumn{1}{|c}{$124$} & \multicolumn{1}{c}{$182$} &
\multicolumn{1}{c}{$0.0$} & \multicolumn{1}{c}{$-1.08$} &
\multicolumn{1}{c}{$0.00$} & \multicolumn{1}{c}{$-1.08$} &
\multicolumn{1}{c}{$-0.08$} & \multicolumn{1}{c}{}%
\end{tabular}
\caption{Equilibrium deformations as well as deformation energies for the
cerium isotopic chain. The columns give successively the number of neutrons
$(N)$, the mass number $(A)$, the prolate equilibrium deformation
$(\beta_{pro}),$ the minimum of the prolate well $(minpro)$, the oblate
equilibrium deformation $(\beta_{obl}),$ the minimum of the oblate well
$(minobl)$, the deformation energy $(E_{def},$see text$),$ the experimental
equilibrium deformation $(\beta_{\exp})$. Note: The deformation energy is
always given for the prolate equilibrium shape because no absolute minimum is
obtained for oblate shape. }%
\label{t2}%
\end{table}

\begin{figure}[ptbh]
\includegraphics[width=160mm,keepaspectratio]{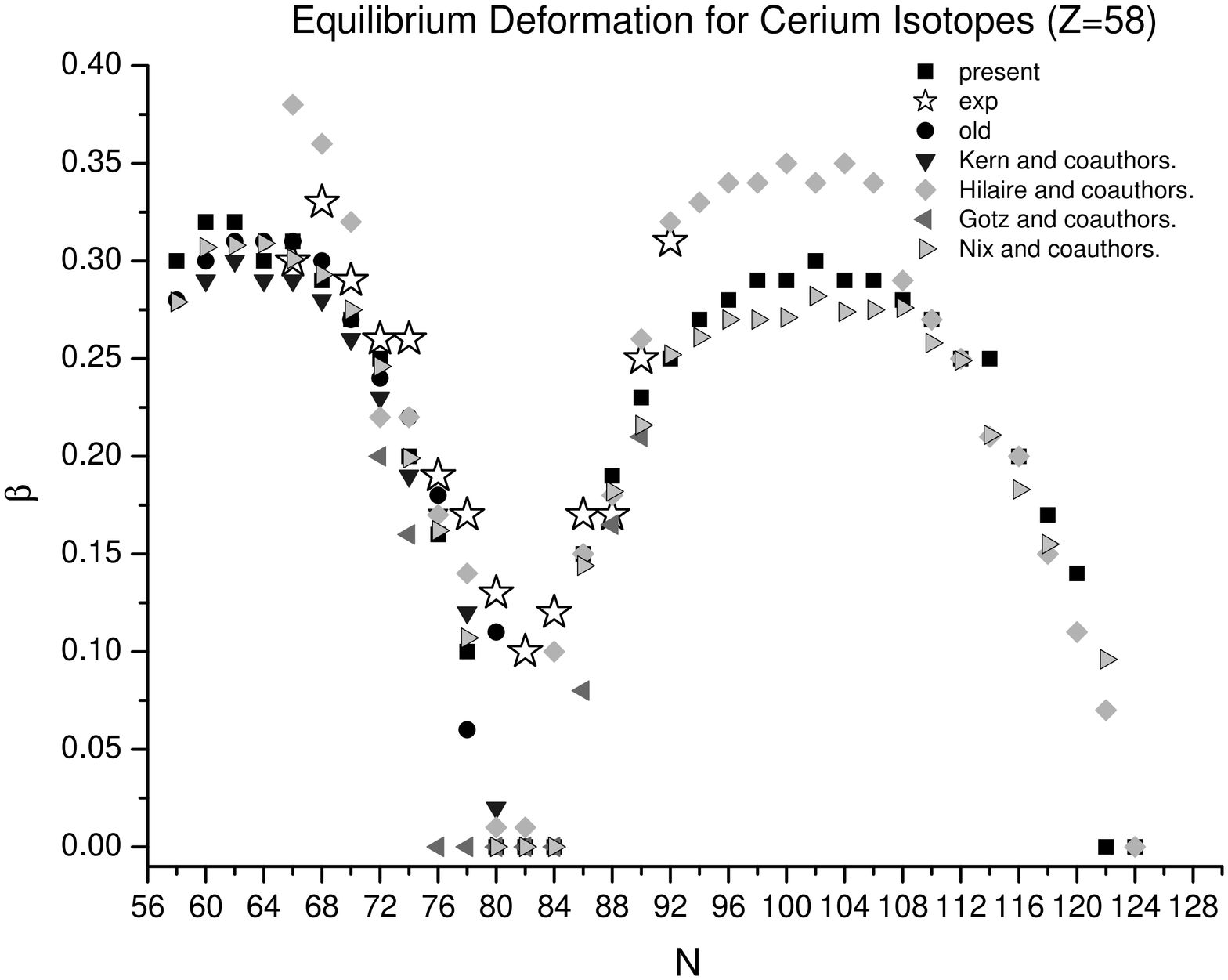}\caption{Theoretical
equilibrium deformations for even-even cerium isotopes evaluated by different
or similar approaches.}%
\label{peok}%
\end{figure}

In Fig.(\ref{peok}) are displayed the present equilibrium deformations,
experimental values \cite{08} , our "old" calculations \cite{014} and other
studies performed by different authors which are: Kern et al.\cite{018},
Hilaire and Girod \cite{021}, Gotz et al.\cite{019} and Nix et al. \cite{020}.
All calculations are based on Macro-Micro method (with different mean fields
or different parameters). Except the one of Ref.\cite{021} which uses
Hartree-Fock-Bogoliubov model with Gogny force.\newline(j) Near magic number
($N=82$) all calculations give spherical equilibrium deformation whereas
experimental results are always slightly deformed (even for $N=82$). It seems
difficult to overcome this defect with a pure static approach which neglects
the role of the mass parameters.\newline(jj) The overall tendency of these
calculations is the same except the fact that HFB calculations differ
significantly from the others with higher values in some regions.\newline(jjj)
Apart from HFB calculations, theoretical values are generally quite close from
each others.\newline(jv) Our old and new calculations give very close results
(see Table \ref{newold}). Thus, even if it is better to choose a proper set of
mean-field parameters for each nucleus, we do not commit a significant error
by taking the same set of parameters for nuclei that do not differ strongly by
the number of neutrons ($N$).

\begin{table}[ptbh]%
\begin{tabular}
[c]{c||ccccccccccccc}%
Cerium $(Z=58)$ & $N=58$ & $60$ & $62$ & $64$ & $66$ & $68$ & $70$ & $72$ &
$74$ & $76$ & $78$ & $80$ & $82$\\\hline\hline
Present $\beta$ & $+0.30$ & $+0.32$ & $+0.32$ & $+0.31$ & $+0.30$ & $+0.29$ &
$+0.27$ & $+0.25$ & $+0.20$ & $+0.16$ & $+0.10$ & $+0.00$ & $+0.00$\\
Old $\beta$ & $+0.28$ & $+0.30$ & $+0.31$ & $+0.31$ & $+0.31$ & $+0.30$ &
$+0.27$ & $+0.24$ & $+0.22$ & $+0.18$ & $+0.06$ & $+0.11$ & $+0.00$\\\hline
Present $E_{def}(MeV)$ & $4.80$ & $5.87$ & $6.19$ & $6.68$ & $6.17$ & $5.43$ &
$4.67$ & $3.34$ & $1.97$ & $0.93$ & $0.19$ & $0.00$ & $0.00$\\
Old $E_{def}(MeV)$ & $4.82$ & $5.77$ & $6.03$ & $6.31$ & $7.08$ & $5.36$ &
$4.41$ & $3.35$ & $2.13$ & $0.77$ & $0.00$ & $0.24$ & $0.00$%
\end{tabular}
\caption{New equilibrium deformations and deformations energies vs old
\cite{014}.}%
\label{newold}%
\end{table}

\subsection{Mass Excesses}

\begin{figure}[ptb]
\includegraphics[width=180mm,keepaspectratio]{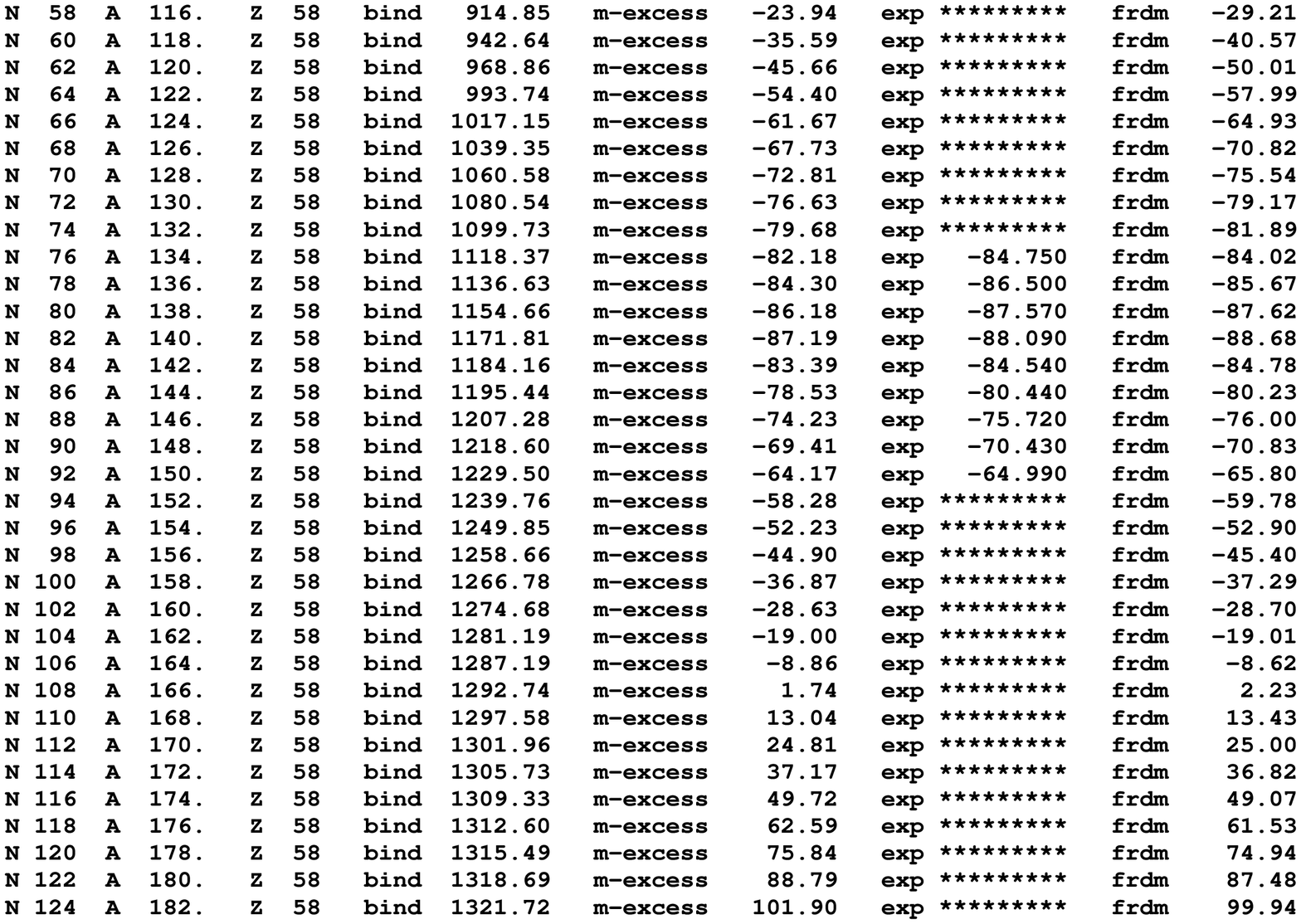}\caption{ Theoretical
binding energies and mass excesses of the present approach compared to the
experimental mass excesses and the ones given by the FRDM model of Ref.
\cite{010}. All energies are expressed in MeV. The experimental data as well
as the frdm results have been entered manually in the code. Asterics mean that
no experimental data is available for the corresponding nucleus.}%
\label{mexc1}%
\end{figure}

We list from a FORTRAN file (see Fig. \ref{mexc1} ) the results of our
theoretical calculations of the binding energies and mass excesses (m-excess)
for the even-even cerium isotopic chain. For the sake of completeness,
experimental mass excesses and the ones of the FRDM model (see Ref.\cite{010})
are also given. We must keep in mind that only $6$ parameters are used in the
liquid drop model whereas $16$ parameters are necessary in the FRDM model.
This explains the "better quality" of the FRDM model. However, we have checked
that the variations of binding energy or mass excesses from one isotope to the
nearest is practically the same in our model and the one of FRDM (the
deviations are about $\pm0.35MeV$). For this reason, the calculation of the
two neutron separation energies (see the following subsection \ref{sub}) will
almost be probably the same for the two approaches even though our model is
not so accurate.

\subsection{Transitional regions in cerium isotopes\label{sub}}

\begin{figure}[ptb]
\includegraphics[angle=90,width=200mm,keepaspectratio]{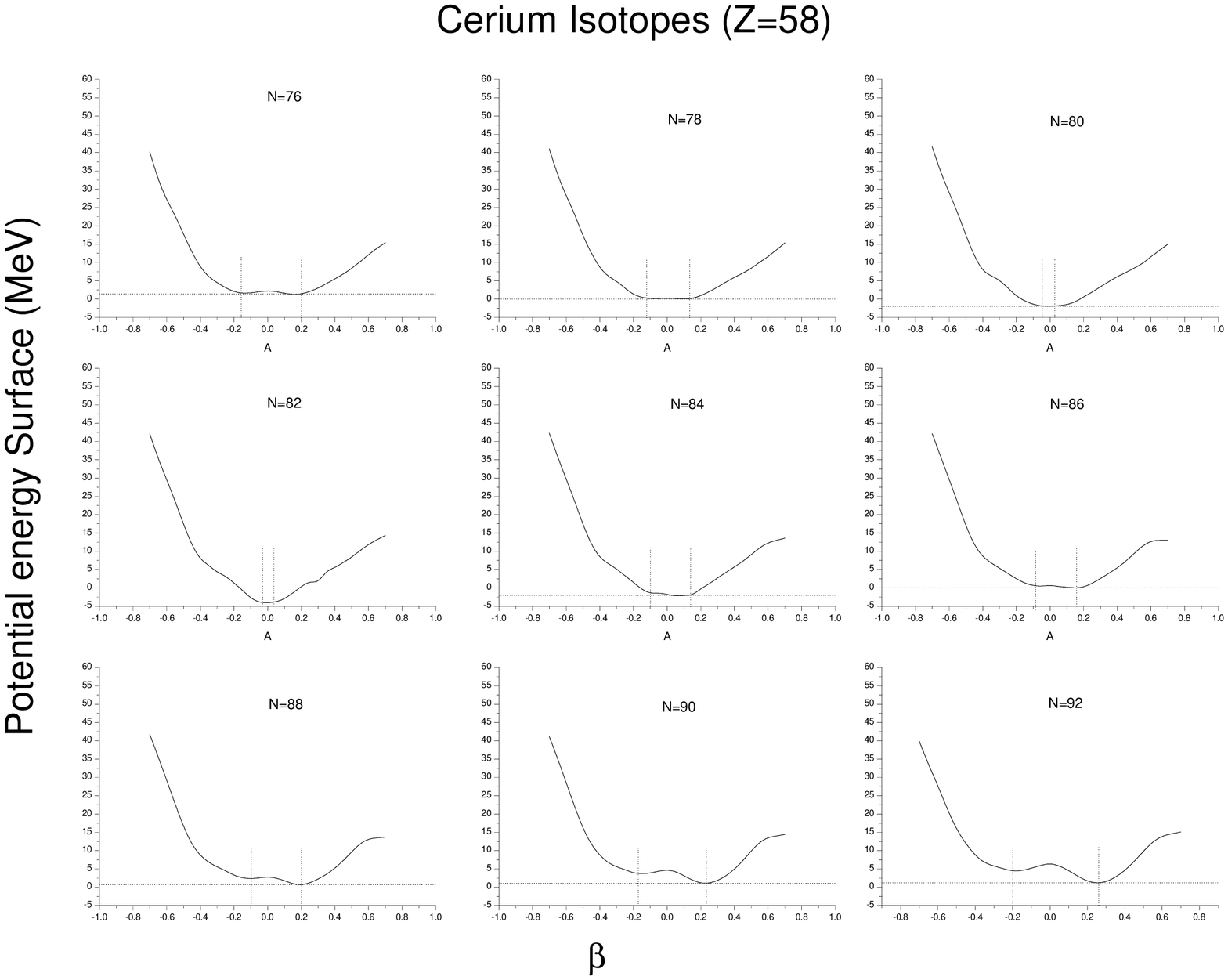}\caption{Shape
evolution for cerium isotopes from $N=78$ to $N=92$.}%
\label{trans}%
\end{figure}

In Fig. \ref{trans} is shown the gradual transition in the potential energy
surface from spherical vibrator to the axially deformed rotor when the number
of neutrons $(N)$ increases from $76$ to $92$. One signature of $X(5)$
symmetry which is a critical-point of phase/shape transitions (quantum phase
transition between spherical and axial symmetries) should be a long flatness
of the potential energy surface with eventually a weak barrier from prolate to
oblate shapes. In this figure, for $N>82$, the width of the flatness increases
as one moves away from $N=82$ but at the same time the difference between
oblate and prolate minima and \ barrier between oblate and prolate shapes also
increase. For example the differences between oblate and prolate energy minima
and barriers for isotopes with $N=88,90,92$ are respectively about $1.5MeV$,
$2.5MeV$ and $3.3MeV$ with energy barrier about $2MeV$, $4MeV$ and $5.5MeV$
respectively. The wideness of the bottom of the well must be relativized with
the height of the barrier. Thus for the case of $N=92$ the width is important,
i.e. about $\Delta\beta\approx\beta_{pro}-\beta_{obl}\approx
0.26-(-0.20)\approx0.46$ but the barrier is about $5.5MeV$ and therefore seems
too high. The case $N=90$ gives a width of $\Delta\beta\approx0.3$ with a
barrier of about $4MeV$. For $N<82$, the case $N=76$ seems to be relatively
equivalent to $N=90$ with a slightly smaller width and a lower height barrier.
Thus it is difficult to determine clearly the existence of a $X(5)$
critical-point. Thus, everything seems to indicate a continuous transition.

\begin{figure}[ptb]
\includegraphics[width=140mm,keepaspectratio]{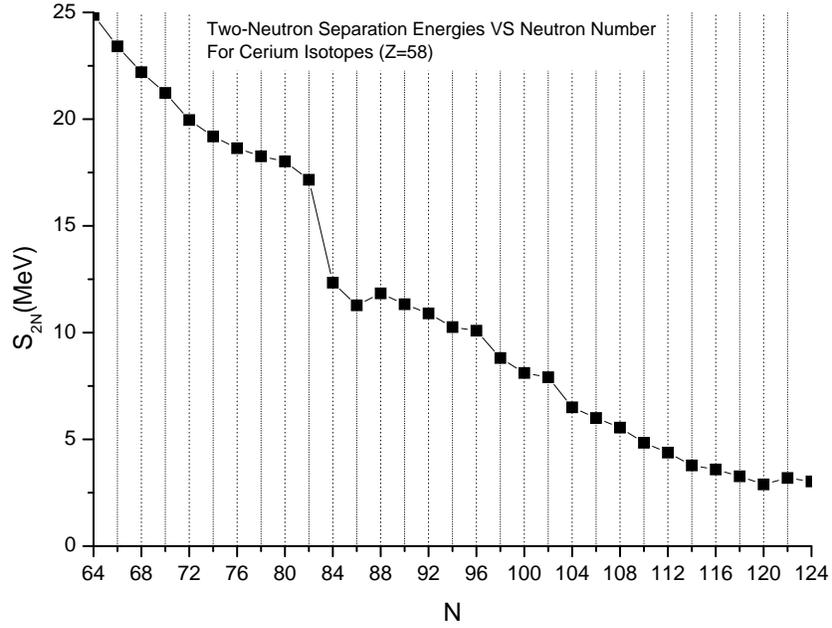}\caption{Two-neutron
separation energies $(S_{2N})$ along the cerium isotopic chain. This quantity
is defined as $S_{2N}(A,Z,N)=Bind(A,Z,N)-Bind(A-2,Z,N-2)$ where the binding
energy $Bind(A,Z,N)$ is given by Eq. (\ref{binding}). Note that in our
approache the neutron drip line (where $S_{2N}\approx0$) can be extrapolated
around $N=128$ for Cerium isotopes.}%
\label{s2n}%
\end{figure}

\begin{figure}[ptb]
\includegraphics[width=100mm,keepaspectratio]{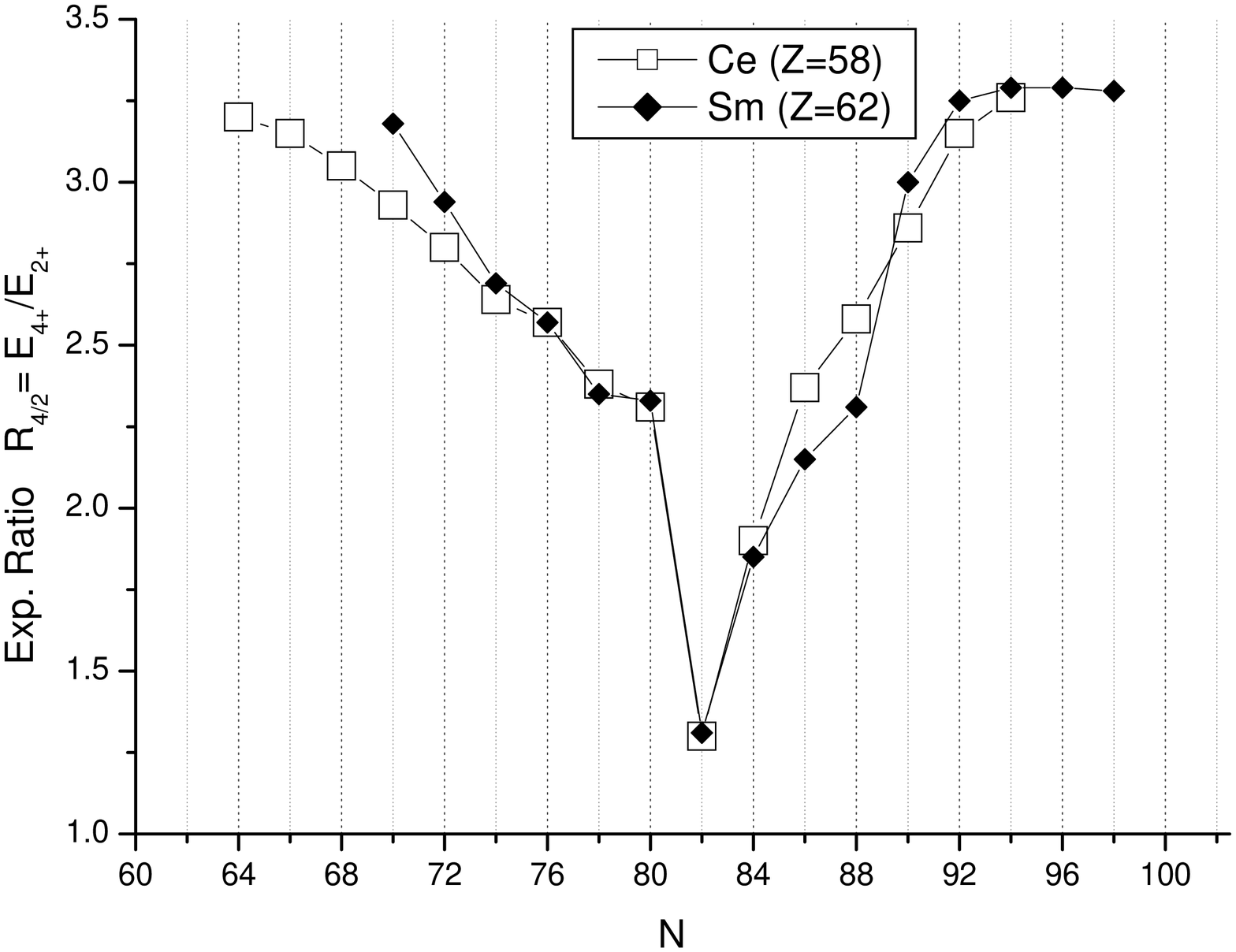}\caption{$R_{4/2}$ enery
ratio as function of neutron number for Cerium and Samarium isotopes. Sudden
variations are associated with magic closure shells for the both chains $($ at
$N=82)$ and with $X(5)$ critical point which occurs only for $Sm$ (at
$N=90$).}%
\label{r42}%
\end{figure}

In Fig. \ref{s2n} is displayed the two-neutron separation energy (TSN) as
function of the neutron number $N$. A clear jump is seen from $N=82$ to
$N=84$, i.e. from one major shell to the following. Just before $N=82$ and
just after $N=84$ the TSN varies more slowly. Far for the "jump" the curve
becomes quasi-linear. Once again, no special behavior is noted around $N=90$
which from Ref.\cite{029} and \cite{030} should constitute \ with $Z\approx62$
the first order shape transition ($X(5)$ critical-point) in the rare earth
region. In Ref.\cite{031}, it has been pointed out that "Empirical evidence of
transitional symmetry at the $X(5)$ critical-point has been observed in
$^{150}Nd$, $^{152}Sm$, $^{154}Gd$, and $^{156}Dy$".\newline One of the most
important signatures of the phase transition is given by a sudden jump in the
value of the energy ratio $R_{4/2}=4_{1}^{+}/2_{1}^{+}$ from one nucleus to
the next. We found it useful to compare the experimental values of this ratio
(see Fig. \ref{r42}) in the cases of the isotopic chains of $Ce$ and $Sm$ (The
experimental values of the considered levels have been deduced from the
adopted level of $ENSDF$ site \cite{032}). The figure shows clearly two facts.
First, the important variation of $R_{4/2}$ near of the magic number $N=82$
for the both isotopic chains and then, the important difference between the
behavior the two isotopic chain from $N=88$ to $N=90$. In effect in the case
of the Samarium, there is a sudden increase of this ratio whereas this is not
the case for the Cerium isotopes. This has been attributed to the $X(5)$
critical-point symmetry of the nucleus $^{152}Sm$. Thus the present study
confirms that cerium isotopic chain is characterized by a continuous
shape/phase transition.

\section{Conclusion}

Potential energy surfaces have been drawn for the cerium isotopic chain. All
even-even nuclei between the two drip lines have been considered. To this end,
we have used the microscopic macroscopic method in which the quantum
corrections have been evaluated by a semi-classical procedure. The microscopic
model is based on a "realistic" Schrodinger equation including a mean field of
a Woods-Saxon type. The macroscopic part of the energy is evaluated from the
liquid drop model using the version of Myers and Swiatecki. The following
points must be remembered:\newline(i) All equilibrium deformations have been
found prolate with an important deformation energy compared to oblate shapes.
\newline(ii) The maximum deformations are of order $\beta\approx0.3$ and are
located around $N=64$ and $N=102$ with deformation energy about $6MeV$ and
$9MeV$ respectively. The equilibrium deformations decrease as one moves away
from these two nuclei.\newline(iii) Spherical shapes are found in the
neighborhood of $N=82$.\newline(iv) Good agreement is obtained between
theoretical and experimental values if one excepts the area of the shell
closure $N=82$ where the latter are slightly larger.\newline(v) This isotopic
chain possesses a continuous shape/phase transition from spherical shapes
toward the axially symmetric ones.

\begin{center}

\end{center}

\appendix

\section{Constants of the binding energy of the liquid drop model}

The constants of Eq.(\ref{binding}) are defined as follows:%

\begin{align*}
C_{V}  &  =a_{V}\left[  1-\kappa I^{2}\right]  \text{ \ (in the volume
term)\ \ \ \ \ \ \ \ \ }\\
\text{\ }C_{S}  &  =a_{S}\left[  1-\kappa I^{2}\right]  \text{ \ (in the
surface term)}\\
\text{\ \ \ \ \ \ \ \ \ }I  &  =\frac{N-Z}{N+Z}\text{ (relative neutron
excess)}\\
\varepsilon &  =+1(even-even),0(odd),-1(odd-odd)\text{ (in the pairing
term)}\\
\text{\ \ \ \ \ \ }C_{C}  &  =\frac{3}{5}\frac{e^{2}}{r_{0}}\text{ \ \ \ (in
the Coulomb term)\ \ \ \ \ }\\
\text{\ }C_{d}  &  =\frac{\pi^{2}}{2}\left(  \frac{a_{0}}{r_{0}}\right)
^{2}\frac{e^{2}}{r_{0}}\text{ (diffuseness correction)}%
\end{align*}
The last correction to the Coulomb energy takes into account that the liquid
drop has not a sharp but a diffuse surface of the Woods-Saxon type [ ]. The
diffuseness parameter is $a_{0}$ and the charge radius "contains" $r_{0}$
($R_{ch}=r_{0}A^{1/3}$).

\section{Constants of the Woods-Saxon mean potential}

"Universal parameters" of the Woods-Saxon central and Spin-orbit potentials
entering in Eq.\ref{pot}.%

\begin{tabular}
[c]{llll}
&  & Neutrons & Protons\\
Central mean field & Depth $(MeV)$ & $V_{0neut}=49.6(1-0.86I)$ &
$V_{0prot}=49.6(1+0.86I)$\\
Central mean field & Radius $(fm)$ & $R_{Vneut}=1.347A^{1/3}$ & $R_{Vprot}%
=1.275A^{1/3}$\\
Spin-orbit mean field & SO-coupling strength $\lambda$ & $35.0$ & $36.0$\\
Spin-orbit mean field & Radius $(fm)$ & $R_{SO-neut}=1.310A^{1/3}$ &
$R_{SO-nprot}=1.200A^{1/3}$\\
Central mean field & diffuseness $(fm)$ & $a_{0}=0.70$ & $a_{0}=0.70$\\
Spin-orbit mean field & diffuseness $(fm)$ & $a_{0}=0.70$ & $a_{0}=0.70$%
\end{tabular}

\end{document}